\documentclass[12pt]{iopart}

\usepackage{graphicx}
\begin{document}

\title{Statistical Model Predictions for p-p and Pb-Pb collisions at LHC}

\author{I Kraus${}^{1,4}$, J Cleymans${}^{2}$, H Oeschler${}^{1}$, K Redlich${}^{3}$, S Wheaton${}^{1,2}$}

\address{   ${}^{1}$Institut f\"ur Kernphysik, Darmstadt University of Technology, D-64289 Darmstadt, Germany\\
		${}^{2}$UCT-CERN Research Centre and Department  of  Physics,\\ University of Cape Town, Rondebosch 7701, South Africa\\
		${}^{3}$Institute of Theoretical Physics, University of Wroc\l aw, Pl-45204 Wroc\l aw, Poland\\
		${}^{4}$Nikhef, Kruislaan 409, 1098 SJ Amsterdam, The Netherlands}

\begin{abstract}
Predictions for particle production at LHC are discussed in the context of the statistical model.
Moreover, the capability of particle ratios to determine the freeze-out point experimentally is studied, and the best suited ratios are specified.
Finally, canonical suppression in p-p collisions at LHC energies is discussed in a cluster framework. Measurements with p-p collisions will allow us to estimate the strangeness correlation volume and to study its evolution over a large range of incident energies.
\end{abstract}

\maketitle

Particle production in heavy-ion collisions is, over a wide energy range, consistent with the assumption that hadrons originate from a thermal source with a given temperature $T$ and a given baryon chemical potential $\mu_B$. In the framework of the statistical model, we exploit the feature that the freeze-out points appear on a common curve in the $T - \mu_B$ plane. The parameterization of this curve, taken from Ref.[1],
is used to extrapolate to the LHC energy of $\sqrt{s_{nn}}$ = 5.5 TeV: $T \approx $170 MeV, $\mu_B \approx$ 1 MeV.

For the given thermal conditions, particle ratios in central Pb-Pb collisions were calculated; numerical values are given in Ref.[2]. As soon as experimental results become available, the extrapolation can be cross-checked with particle ratios that exhibit a large sensitivity to the thermal parameters. The ratios shown in figure 1 (left) hardly vary over a broad range of $T$ and $\mu_B$. This feature can be used to investigate the validity of the statistical model at LHC: Especially the prediction for the $K/\pi$ ratio is limited to a narrow range. It would be hard to reconcile experimental results outside of this band with the statistical model.
\\
Antiparticle over particle ratios, on the other hand, strongly depend on $\mu_B$ (figure 1 middle). Most of all, the $\bar{p}/p$ ratio almost directly translates to the baryon chemical potential, since the $T$ dependence is very weak. Better suited for the temperature determination are ratios with large mass differences, i.e. $\Omega/\pi$ and $\Omega/K$, which increase in the studied range by 25\% per 10 MeV change in $T$. 
The astonishing similarity between $K$ and $\pi$ in this respect is caused by the huge contribution of 75\% from resonance decays to pions for the given thermal conditions (Ref.[3]).

\begin{figure}[htb]
\begin{minipage}[b]{0.29\linewidth}
\centering
\includegraphics*[width=\linewidth]{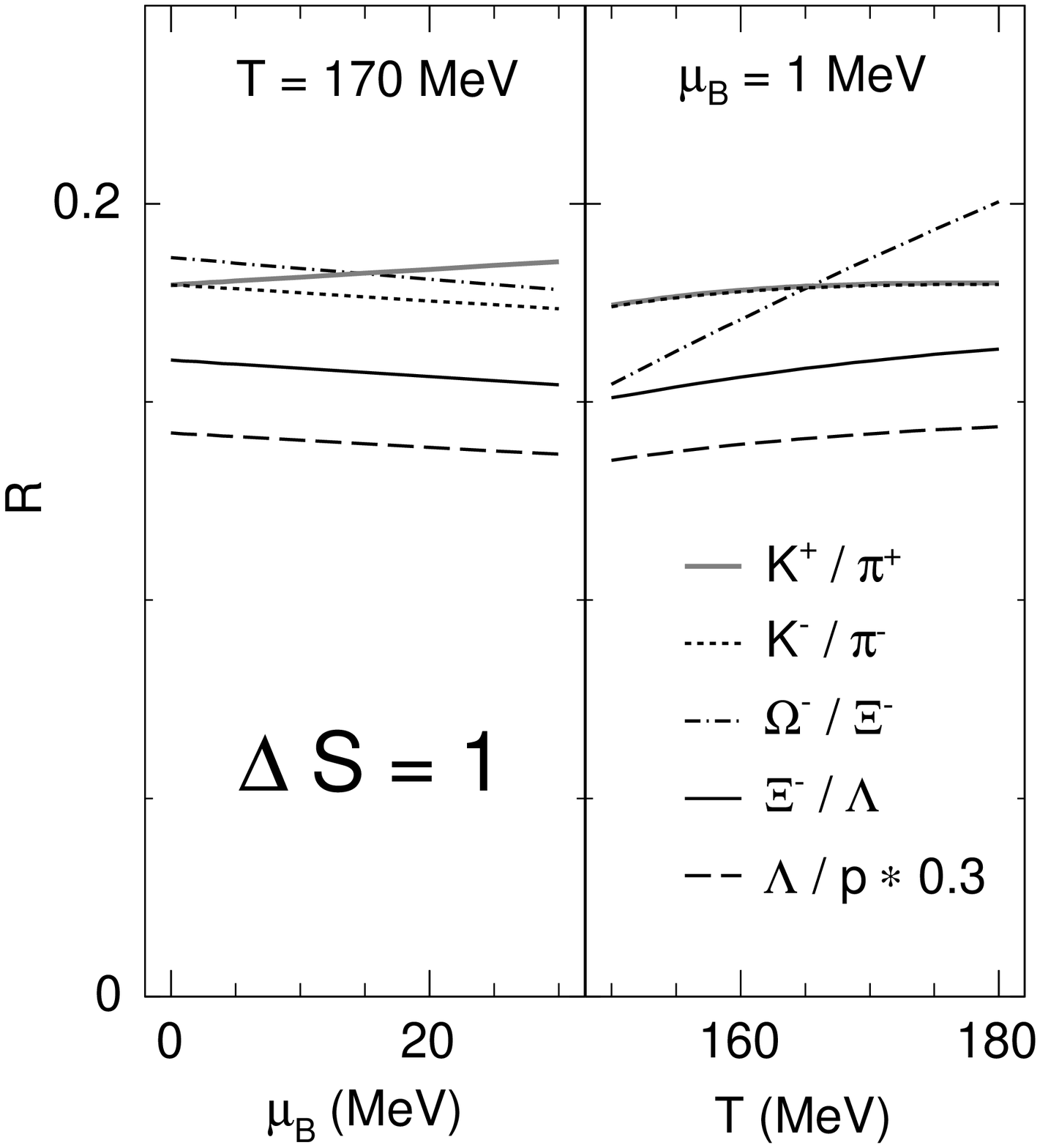}
\end{minipage}\hfill
\begin{minipage}[b]{0.36\linewidth}
\centering
\includegraphics[width=\linewidth]{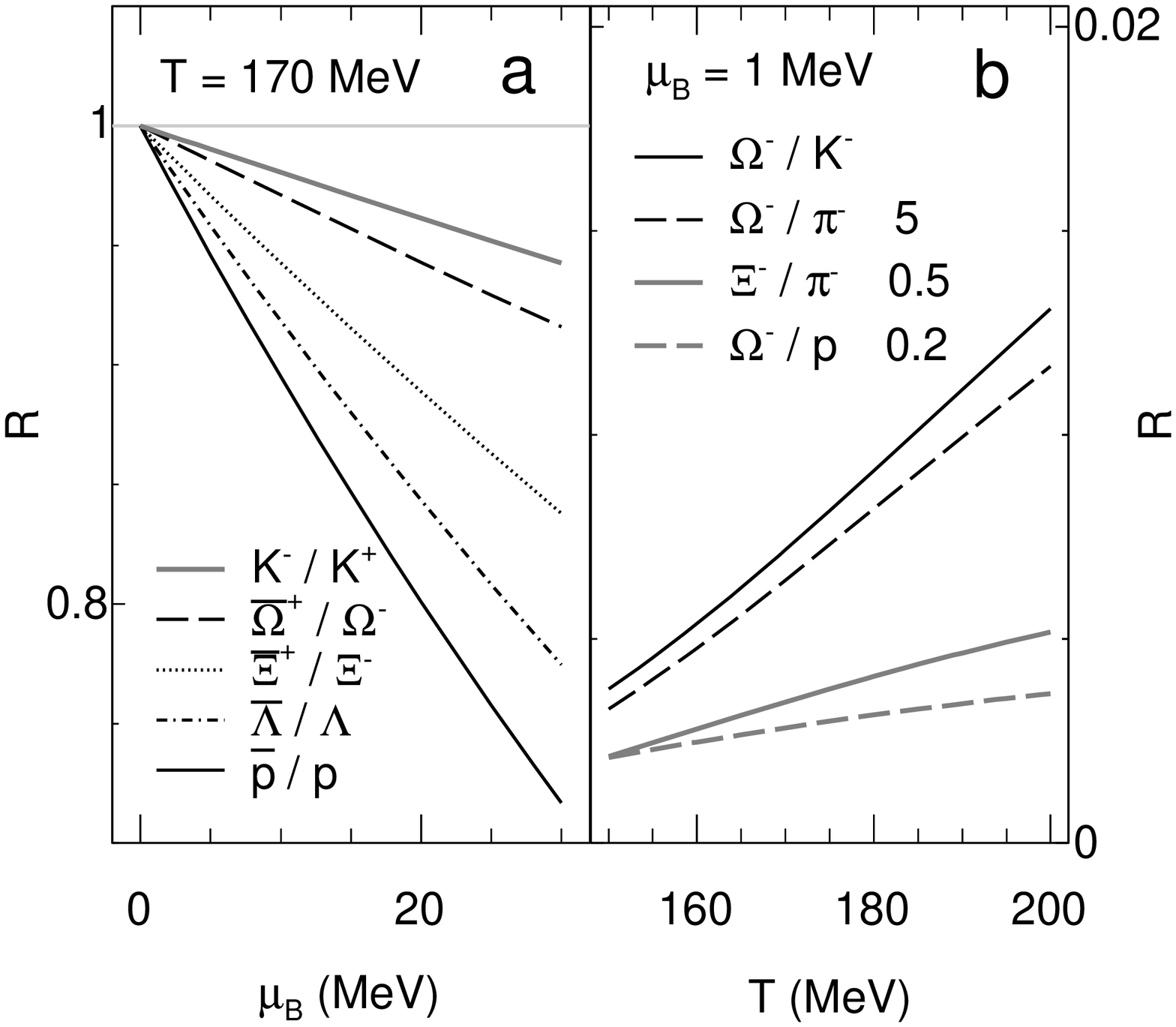}
\end{minipage}
\begin{minipage}[b]{0.34\linewidth}
\centering
\includegraphics*[width=\linewidth]{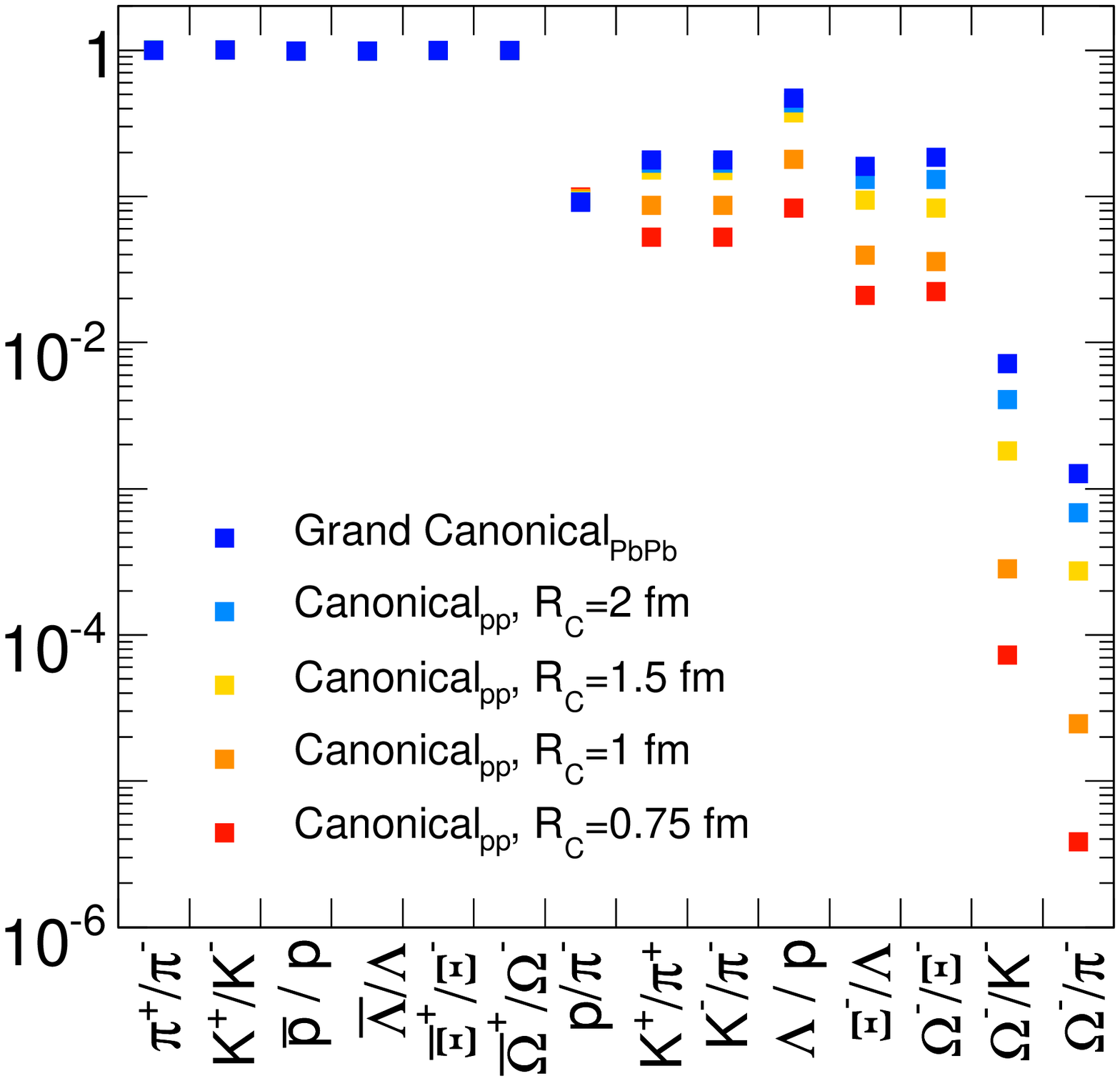}
\end{minipage}\hfill
\caption{\label{fig}
Left: Ratios $R$ of particles with unequal strangeness content as a function of $\mu_B$ for $T$ = 170 MeV (left) and as a function of $T$ for $\mu_B$ = 1 MeV (right).
\\
Middle: Antiparticle/particle ratios $R$ as a function of $\mu_B$ for $T$ = 170~MeV (left)
		(the horizontal line at 1 is meant to guide the eye).
		Particle ratios $R$ involving hyperons as a function of  $T$ for $\mu_B$ = 1 MeV (right).
\\
Right: Ratios $R$ of particles in the grand-canonical ensemble and with suppressed strange-particle phase-space in different canonical volumes indicated by the spherical radius $R_C$, calculated at $\mu_B$ = 1 MeV and $T$ = 170 MeV.} 
\end{figure}

In collisions of smaller systems, the strange-particle phase-space exhibits a suppression beyond the expected canonical suppression. A modification of the statistical model is proposed in Ref.[4], which is based on the assumption that strangeness conservation is maintained in correlated sub-volumes of the fireball. The size of these clusters, which could be smaller than the volume defined by all hadrons, was estimated from relative strangeness production in collisions of small systems at top SPS and RHIC energy. The radius $R_C$ of a spherical cluster is of the order of 1~-~2~fm and shows only a weak energy dependence.
Additionally it is not clear at which stage of the interaction the strangeness abundance is formed. Possibly the early, dense phase is crucial, so the cluster size should be the same at RHIC and LHC, or, on the contrary, the total number of particles at the late stage of hadronisation is relevant; thus $R_C$ should increase as the multiplicity will increase with colliding energy.
\\
Instead of precise predictions as shown for Pb-Pb collisions, the correlation volume will be extracted from measurement. As displayed in figure 1(right), especially the $\Omega/\pi$ ratio varies over orders of magnitude in a reasonable range of the correlation length. This allows for a good estimate of the cluster size which will give us more insight into the mechanism of strangeness production.

\vspace{2mm}

\numrefs{1}
\item Cleymans J, Oeschler H, Redlich K, Wheaton S 2006 {\it Phys. Rev.} C{\bf 73} 034905
\item Cleymans J, Kraus I, Oeschler H, Redlich K, Wheaton S 2006 {\it Phys. Rev.} C{\bf 74} 034903
\item Kraus I, Cleymans J, Oeschler H, Redlich K, Wheaton S 2006 {\it J. Phys.} G{\bf 32} S495	
\item Kraus I, Oeschler H, Redlich K 2005 {\it PoS(HEP2005)140}; detailed publication in preparation

\endnumrefs

\end{document}